\magnification=1200  
\vskip .26in
\hskip 1.0in%

{\vbox{\centerline{\bf A (1,2) Heterotic String  }
\centerline{\bf with Gauge Symmetry }}

\smallskip
\centerline{David M. Pierce}
\bigskip\centerline{\it Institute of Field Physics}
\centerline{\it Department of Physics and Astronomy}
\centerline{\it University of North Carolina}
\centerline{\it Chapel Hill, NC 27599-3255, USA}
\bigskip

We construct a (1,2) heterotic string with gauge symmetry and 
determine its particle spectrum. This theory has a local N=1 
worldsheet supersymmetry for left movers and a local N=2 worldsheet
 supersymmetry for right movers and describes particles in either two
 or three space-time dimensions.  We show that fermionizing the 
bosons of the compactified N=1 space leads to a particle spectrum 
which has nonabelian gauge symmetry.  The  fermionic formulation of
 the theory corresponds to a dimensional reduction of self dual Yang
 Mills.  We also give a worldsheet action for the theory and 
calculate the one-loop path integral.

\vskip10pt
\centerline{\bf I. Introduction}
\vskip10pt
 N=2 string theories were first studied by Ademollo et al.[1]. They
 constructed a string which has a local N=2 worldsheet supersymmetry and found that the critical dimension was D=2. They also found that
 the  only physical state was the ground state. Some years later, it
 was discovered that the D=2 dimensions were complex, implying that
 the theory actually exists in 4 space-time dimensions [2,3]. As a 
consequence, the metric of the space is (2,2) or (0,4). More 
recently, heterotic constructions of N=2 theories have been 
proposed[4,5]. The (0,2) heterotic theory, which is a combination of
 the N=0 bosonic string and the N=2 string, contains a number of 
massless scalars in two or three space-time dimensions. Compactifying
 on an even self dual lattice results in states which transform in 
the adjoint representation of a gauge group. Furthermore, the two 
dimensional version of the theory is equivalent to a dimensional 
reduction of four-dimensional self dual Yang Mills. When using the 
same compactification scheme for (1,2) theories, that is theories 
with N=1 local worldsheet supersymmetry for the left movers and N=2
 worldsheet supersymmetry for the right movers, the physical states 
are again scalars in 2 or 3 space-time dimensions but they do not 
have group quantum numbers. This leads us to question whether there 
are other compactifications with massless particles in a non-trivial 
representation of a gauge group. This extends the 
general picture of the gauge symmetries which couple to critical strings. 

In this paper we show that the (1,2) heterotic string can support non-abelian gauge symmetry.  In sect. II, we discuss the action for the (1,2) theory and derive the classical equations of motion. In sect. III we derive the BRST charges and the anomaly free algebra.  In sect. IV, we discuss the physical spectrum of the (1,2) string. In the fermionic case, the states form representations of any 24 dimensional gauge group, and this is discussed in sect. V.  In sect. VI, we calculate the partition functions for the two compactifications of the (1,2) heterotic theory. Finally, in sect. VII, we discuss the correspondence of the  fermionic theory to self-dual Yang Mills. 
\vskip10pt

\centerline{\bf II Action for (1,2) Theory}
\vskip10pt
We would like to construct an action which has local N=1 worldsheet supersymmetry for left movers and local N=2 supersymmetry for the right movers. The conventional N=2 action is given by[1,2,3]:

$$\eqalign{S_1=&-{1\over {2\pi}}\int d^2\sigma e\{h^{\alpha \beta}\partial_{\alpha}z^{\gamma}\partial_{\beta}\bar z_{\gamma}-i\bar \psi ^{\gamma}\rho ^{\alpha}{\textstyle\partial }^{\hskip-6pt{}^\leftrightarrow}_{\alpha}\psi_{\gamma}+A_{\alpha}\bar \psi^{\gamma}\rho^{\alpha}\psi_{\gamma}+\cr & (\partial_{\alpha}z^{\gamma}-{1\over 2}\bar \chi_{\alpha}\psi^{\gamma})\bar \psi_{\gamma}\rho^{\beta}\rho^{\alpha}\chi_{\beta}+(\partial_{\alpha}\bar z^{\gamma}-{1\over 2}\bar \psi^{\gamma}\chi_{\alpha})\bar \chi_{\beta}\rho^{\alpha}\rho^{\beta}\psi_{\gamma}\}}\eqno(2.1)$$
where $z^{\gamma}(\sigma,\tau)$ is a complex bosonic field valued on the worldsheet; $z^{\gamma}=X^{2\gamma}+iX^{2\gamma+1}$ for $1\le \gamma \le 2$. $\psi^{\gamma}(\sigma,\tau)$ is a complex fermion on the worldsheet, $A_{\alpha}$ is a gauge field which is required to make the U(1) symmetry local, and the complex gravitino, $\chi_{\alpha}$, is required to make the N=2 supersymmetry local. 
To restrict the N=2 susy to the right movers, we keep only 1 spinor component of the fermions and gravitino. They satisfy $\rho^+\psi=0$ and  $\rho^-\chi_{\alpha}=0$ where $\rho^{\alpha}$ are the two-dimensional gamma matrices.  The N=2 susy transformations are :
$$\delta z={\bar \epsilon}_{r} \psi$$
$$\delta \psi=-i\rho^{\alpha}\epsilon_{r}\partial_{\alpha}z\eqno(2.2)$$
The spinor parameter satisfies $\rho^-\epsilon_{r} =0$.

One can not take the left movers to be the standard fermionic string[5]. The metric of the N=1 string is (9,1) and the metric for the N=2 string is (2,2) so there is no common subspace. The solution to this problem is found by recognizing that the bosonic fields of the N=2 supergravity multiplet when gauged do not refer to the left or right movers. This means that the two-dimensional graviton field and the Maxwell field of the N=2 supergravity multiplet is common between left and right movers.  It is necessary to gauge a U(1) current for the left movers. This current introduces  additional ghosts-antighosts with conformal weights (1/2,1/2) and (1,0). This changes the ghost contribution to the anomaly to $-1-2-15= -18$ so that the critical dimension is now 12. The new ghost system brings an additional time component so the metric is (2,10)[6]. 

The N=1 action has the form:

$$S_2=-{1\over {2\pi}}\int d^2\sigma e\{h^{\alpha \beta}\partial_{\alpha}X^{i}\partial_{\beta}X_{i}+\lambda(\partial_{+}X^i)^2-i\bar \psi'^{\mu}\rho^{\alpha}\partial_{\alpha}\psi'_{\mu}\}$$
$$S_3=-{1\over {\pi}}\int d^2\sigma e \bar\chi^{\prime}_{\alpha}\rho^{\beta}\rho^{\alpha}\psi'^{\mu}\partial_{\beta}X_{\mu}$$
$$S_4=-{1\over {4\pi}}\int d^2\sigma e \bar \psi'_{\mu}\psi'^{\mu}\rho^{\beta}\rho^{\alpha}\chi^{\prime}_{\beta}$$
Here, i is an index for the internal space, and $\mu$ runs over all space.
The N=1 susy is restricted to the left movers by keeping only the left handed fermions; $\rho^-\psi'=0$ and $\rho^+\chi^{\prime}=0$. N=1 worldsheet supersymmetry for the left movers also requires adding a superpartner for the gauge field, $A_{\alpha}$, which is denoted by $\psi^{\prime\prime}$.  We also have the restriction $\rho^-\psi^{\prime\prime}=0$. The equations of motion for the U(1) gauge fields lead to the constraints   Altogether, the complete list of local symmetries [1] is sufficient to gauge away all components of $A_{\alpha}, \chi_{\alpha}, \chi'_{\alpha}$, and $\psi^{\prime\prime}$. As is customary in heterotic theories, to eliminate the internal bosons $X^i$ for the right moving N=2 part, we add a lagrange multiplier, $\lambda(\partial_{+}X^i)^2$,in the second term[7]. 

This part of the action is invariant under the local N=1 worldsheet susy given by
$$\delta X^{\mu}=\bar \epsilon \psi'^{\mu}$$
$$\delta\psi'^{\mu}=-i\rho^{\alpha}\epsilon(\partial_{\alpha}X^{\mu}-\bar \psi^{\mu}\chi^{\prime}_{\alpha})$$
$$\delta e^{a}_{\alpha}=-2i\bar \epsilon \rho^a\chi^{\prime}_{\alpha}$$
$$\delta \chi^{\prime}_{\alpha}=\partial_{\alpha}\epsilon\eqno(2.4)$$
The spinor parameter, $\epsilon$, satisfies $\rho^+\epsilon=0$.
Here, we have used the usual convention for light cone coordinates, $\sigma^{\pm}=\sigma^0\pm\sigma^1$. 

The U(1) gauge symmetry is:

$$\delta \psi =i\Sigma(\sigma,\tau)\psi\hskip40pt \delta A_{\alpha}=\partial_{\alpha}\Sigma(\sigma,\tau)$$
$$\delta\chi_{\alpha}=i\Sigma(\sigma,\tau)\chi_{\alpha}\hskip40pt \delta e^a_{\alpha}=\delta Z^{\mu}=0\eqno(2.5)$$

In addition to the usual constraints derived from the equations of motion for the metric and gravitino of the N=1 string(Virasoro and superVirasoro currents),  there will be constraints following from the equations of motion for the gauge field $A_{\alpha}$ and its superpartner $\psi^{\prime\prime}$. These additional constraints on the left movers are:
$$J^1(z)=\nu_{\mu}\partial x^{\mu}(z)=0\eqno(2.6)$$
where $\nu$ is a 12 dimensional vector, and 
$$J^2(z)=\nu_{\mu}\psi^{\mu}(z)=0\eqno(2.7)$$

In order for the BRST operator to square to zero, it is necessary that $\nu$ be nilpotent, i.e. $\nu^2=0$. This is also necessary for the anomaly to cancel in the constraint algebra (3.8).

\vskip20pt

\centerline{\bf III. Currents and anomaly free algebra}
\vskip10pt
Following the approach used by Fradkin[8], we determine the BRST charge and constraint equations. This procedure has already been worked out for the N=2 string and is shown in[3]. 
We first give the results for the N=2 string. 
\vskip10pt
The conserved currents are found from the equations of motion for the gauge fields $e_{\alpha}^a$, $\chi$, $\bar \chi$, and $A_{\alpha}$.
We introduce a pair of canonically conjugate ghost oscillators for each
constraint, with statistics opposite to the respective constraint. The conformal weights of the ghosts are given below:

$$L_n: \hskip20pt \{c_m,\bar c_n\}=\delta_{m+n};\hskip10pt (2,-1) $$
$$G_p: \hskip20pt [e_p,\bar e_q]=\delta_{p+q};\hskip10pt (3/2,-1/2)$$
$$\bar G_p:\hskip20pt [f_p,\bar f_q]=\delta_{p+q};\hskip10pt(3/2,-1/2)$$
$$T_m:  \hskip20pt \{t_m,\bar t_n\}=\delta_{m+n};\hskip 10pt(1,0)\eqno(3.1)$$

The anomaly free currents are:
$$\eqalign{\tilde L_m=&\{Q,\bar c_m\}=:L_m+(m-n)\bar c_{m+n}c_{-n}+(1/2m-p)(\bar e_{m+p}e_{-p}+\bar f_{m+p}f_{-p})-\cr
 & n\bar t_{m+n}t_{-n}-A\delta_{m,0}}$$
$$\tilde G_{p}=:G_p-2\bar c_{p+q}f_{-q}-2(p-q)\bar t_{p+q}f_{-q}+(1/2m-p)\bar e_{m+p}c_{-m}+1/2\bar e_{m+p}t_{-m}:$$
$$\tilde {\bar G}=:\bar G -2\bar c_{p+q}e_{-q}+2(p-q)\bar t_{p+q}e_{-q}+(1/2m-p)\bar f_{m+p}c_{-m}-1/2 \bar f_{m+p}t_{-m}:$$
$$\tilde J_m=:J_m-m\bar t_{m+n}c_{-n}+1/2\bar e_{m+p}e_{-p}-1/2\bar f_{m+p}f_{-p}-B\delta_{m,0}:\eqno(3.2)$$
A and B are normal ordering constants.
Note that $f_p=e_{-p}^{\dagger}$ and $\bar f_p=-\bar e_{-p}^{\dagger}$.
The constraints following from the action are:
$$L_n=\sum_{r=Z}\bar \alpha_{n-r}\cdot \alpha_r-\sum_{s=Z-b}[n/2+s]\bar d_{n-s}\cdot d_s\hskip10pt (n\in Z)$$
$$G_p=i\sqrt{2}\sum_{s\in Z-b}\bar \alpha_{p-s}\cdot d_{s}\hskip10pt {p\in Z-b}$$
$$\bar G_{p}=-i\sqrt{2}\sum_{r\in Z}\bar d_{p-r}\cdot \alpha_{r}\hskip10pt {p\in Z+b}$$
$$T_n=-1/2\sum_{s\in Z}\bar d_{n-s}\cdot d_{s}\hskip10pt (n\in Z)\eqno(3.3)$$
where b is the boundary condition on the worldsheet fermions.

We now determine the BRST charge and associated currents for the left moving N=1 part. The constraints or currents follow from the field equations for the respective gauge fields, $e_{\alpha}^a$, $\chi_{\alpha}$, $A_{\alpha}$, and $\psi^{\prime\prime}$. Following the BRST approach, we introduce ghosts which have opposite statistics to every current. The currents following from the action are:

$$L_m=1/2\sum :\alpha_{n}\cdot\alpha_{m+n}:+1/2\sum(r+m/2):b_{r}\cdot b_{m+r}:$$
$$G_r=\sum\alpha_{-n}\cdot b_{r+n}$$
$$J^1_m=\nu\cdot\alpha_{m}$$
$$J^2_r=\nu\cdot b_r\eqno(3.4)$$

The ghosts (and conformal weights) associated with these currents are:  

$$L_n: \hskip20pt \{c_m,\bar c_n\}=\delta_{m+n};\hskip10pt (2,-1)$$
$$G_p: \hskip20pt [e_p,\bar e_q]=\delta_{p+q};\hskip10pt(3/2,-1/2)$$
$$J^2_p:\hskip20pt [f_p,\bar f_q]=\delta_{p+q};\hskip10pt(1/2,1/2)$$
$$J^1_m:\hskip20pt \{t_m,\bar t_n\}=\delta_{m+n};\hskip10pt(1,0)\eqno(3.5)$$

The BRST charge is:
$$\eqalign{Q=&:c_{-n}L_{n}+e_{-p}G_{p}+f_{-p}J^2_{p}+t_{-m}J^1_{m}-{1\over 2}(m-n)c_{-m}c_{-n}\bar c_{m+n}\cr
 & -2e_{-p}e_{-q}\bar c_{p+q}+mc_{-n}t_{-m}\bar t_{n+m}-(n/2+p)c_{-n}f_{-p}\bar f_{n+p}+nt_{-n}e_{-p}\bar f_{n+p}\cr
  &+ f_{-q}e_{-p}\bar t_{q+p}+(1/2m-r)c_{-m}e_{-r}\bar e_{m+r}-ac_{0}-bt_{0}-cf_{0}:}\eqno(3.6)$$
where a,b, and c are normal ordering constants.

The anomaly free currents can now be determined and they are:
$$\eqalign{\tilde L_m=&\{Q,\bar c_m\}=:L_m+(m-n)\bar c_{m+n}c_{-n}+(n/2+p)\bar f_{n+p}f_{-p}\cr &-m\bar t_{n+m}t_{-n}+(1/2m-r)\bar e_{m+r}e_{-r}:-a\delta_{m,0}}$$
$$\tilde J^1_{n}=\{Q,\bar t_n\}=:J^1-n\bar t_{n+m}c_{-m}-n\bar f_{n+p}e_{-p}:-b\delta_{m,0}$$
$$\tilde J^2_{p}=[Q,\bar f_{p}]=:J^2_{p}-(n/2+p)\bar f_{n+p}c_{-n}+\bar t_{p+q}e_{-q}:-c\delta_{p,0}$$
$$\tilde G_{p}=[Q,\bar e_{p}]=:G_p-2\bar c_{p+q}e_{-q}+n\bar f_{n+p}e_{-p}+\bar t_{q+p}f_{-q}+(1/2m-p)\bar e_{m+p}e_{-m}:\eqno(3.7)$$
These currents satisfy the following algebra:
$$[\tilde L_n,\tilde L_m]=(n-m)\tilde L_{n+m}
\hskip40pt[\tilde L_m,\tilde G_r]=(1/2m-r)\tilde G_{m+r}$$
$$[\tilde L_n,\tilde J^1_m]=-m\tilde J^1_{n+m}\hskip40pt
[\tilde L_n,\tilde J^2_{m}]=({n\over 2} +m)\tilde J^2_{n+m}$$

$$\{\tilde G_r,\tilde G_s\}=2\tilde L_{r+s}
\hskip20pt[\tilde J^1_{n},\tilde G_p]=-n\tilde J^2_{n+p}\hskip20pt
\{\tilde J^2_n,\tilde G_p\}=-\tilde J^1_{n+p}$$
$$[\tilde J^1,\tilde J^1]=0\hskip20pt
\{\tilde J^2,\tilde J^2\}=0
\hskip20pt[\tilde J^1,\tilde J^2]=0\eqno(3.8)$$
The nilpotence of the quantum operator Q is necessary and sufficient for the vanishing of the anomaly in the $\tilde K$ algebra. It is also true that the vanishing of the anomaly in the $\tilde K$ algebra implies the nilpotence of the quantum operator Q. $Q^2=0$ is necessary for the consistency of the theory. In practice, it turns out to be simpler to compute the anomaly in $\tilde K$ and require it to vanish than the nilpotence of Q.

We now determine the dimension which is necessary for the vanishing of the anomalies in the constraint algebra and the normal ordering constants in (3.7). For d bosonic and d fermionic fields, the normal ordering constants are:

\noindent Neveu-Schwarz (NS) space-time (antiperiodic supercurrent)
$$a=1/2(1-d_R/16)\eqno(3.9)$$
\noindent Ramond (R) space-time (periodic supercurrent)
$$a=1/2(2/3-d_R/12 + d_{NS}/24)\eqno(3.10)$$
and $d_R$ is the number of internal Ramond fermions and $d_{NS}$ is the number of Neveu-Schwarz fermions. In the particular case when each sector contains fermions of one boundary condition, we see that the critical dimension is 12; a=1/2 when there are 12 NS; and a=0 for when there are 12 R.
When the internal bosons have been fermionized,  the normal ordering constant,a, in the NS or R sectors is given by: 
$$a=1/2(d_{NS}/24-2d_R/48)\eqno(3.11)$$

In the particular case when the fermions in a sector have the same boundary conditions, the critical dimension is D=28(4 space-time fermions + 24 internal fermions(8x3))
a=1/2 when all 28 fermions have NS boundary conditions, and $a= -1$ when all 28 fermions have R boundary conditions.
 
The normal ordering constants, b, and c can be determined from $[\tilde L_m,\tilde J^1_n]$ and $[\tilde L_m,\tilde J^2_p]$.  Since there are no constant singularity terms in the operator
products of $L(z)J(w)$, b and c must be zero.
\vskip20pt

\centerline{\bf IV. Physical States for (1,2) heterotic string}

\vskip10pt
The physical spectrum of the N=2 heterotic theories are determined
by taking the tensor product of the N=0 or N=1 left moving part with the right moving N=2 part. Since all the physical states of the N=2 string are massless, only massless states of the left moving  N=1 part will be physical. In addition, there are further constraints coming from the gauged U(1) current.
\vskip10pt
 
For the N=2 right movers we define the Fock vacuum by:
$$\bar \alpha^{R\mu}_n,\alpha^{R\mu}_n,d^{R\mu}_p,\bar d^{R\mu}_p,c^{R}_n,\bar c^{R}_n,e^{R}_p,\bar e^{R}_p,f^{R}_p,\bar f^{R}_p,t^{R}_n,\bar t^{R}_n |0,k>=0$$
$$(n\in Z,p\in n+\phi-1/2,n>0,p>0); 1\le \mu\le 2$$
$$\bar c^R_0,|0,k>=\bar t^R_0|0,k>=0\eqno(4.1)$$

For the left moving N=1 part:

$$\alpha^{Li}_m,d^{Li}_r,c^{L}_m,\bar c^{L}_m,e^{L}_q,\bar e^{L}_q,f^{L}_q,\bar f^{L}_q, t^L_m,\bar t^L_m|0,k>=0$$
$$(m\in Z;q,r\in n+\phi-1/2,m>0,q>0,r>0); 1\le i\le 4    $$
$$\bar c^L_0,|0,k>=\bar t^L_0|0,k>=0\eqno(4.2)$$
When p or q is an integer, we have
$$\bar e^L_0,|0,k>=\bar f^L_0|0,k>=0$$
$$\bar e^L_0,|0,k>=\bar f^L_0|0,k>=0\eqno(4.3)$$

There is a vacuum doubling which arises from the ghost zero modes. This is resolved by requiring all physical states $|\phi>$ to satisfy :
$$\bar c^R_0|\phi>=\bar t^R_0|\phi>=0$$
$$\bar c^L_0|\phi>=\bar t^L_0|\phi>=0\eqno(4.4)$$

The states of the N=2 string have been analyzed in [9].  If the real bosonic coordinates are required to be periodic when $\sigma$ changes by $2\pi$, then all the states are space-time bosons. There are two sectors:

$$I.\hskip 10pt \psi^{\mu}(\sigma+2\pi)=\psi^{\mu}(\sigma), x^{\mu}(\sigma+2\pi)=x^{\mu}(\sigma)$$
$$II. \hskip10pt\psi^{\mu}(\sigma+2\pi)=-\psi^{\mu}(\sigma),\hskip10pt x^{\mu}(\sigma+2\pi)=x^{\mu}(\sigma)\eqno(4.5)$$
The physical states of the theory are required to satisfy $Q|\phi>=0$ where Q is the BRST charge and where $|\phi>$ is a state which does not satisfy $|\phi>=Q|\chi>$.  
 The only physical state in sector I is the massless scalar  ground state $|0,p>$. In sector two, the only physical state is the massless ground state $p^*_{\mu}d^{\mu}_0|0,p>$. This vacuum also is defined by $\bar d^{\mu}_0|0,p>=0$. The second sector, which would in the N=1 string be a space-time fermion, is a space-time boson in the N=2 string. BRST invariance requires that the U(1) current annihilate  all physical states. So we see that the U(1) current excludes the existence of space-time fermions in the theory. 

 For the left moving N=1 part, the solutions can be expressed in terms of complex oscillators. By the same reasoning used for the N=2 string, it can be shown that both the NS and R sectors are space-time bosons. When the internal bosons are fermionized, all the physical particles occur in the NS sector.
 
\vskip10pt
BRST invariance of states with oscillators acting on the vacuum requires that $\epsilon\cdot k=0$.
This is the usual constraint found in any string theory and amounts to keeping
d-2 polarizations. In addition, we have constraints from the U(1) currents. There is the condition of being annihilated by $J^1_m$ for $m\ge 0$. For m=0, we have 

$$\tilde J^1_0|k,*>=\nu\cdot k|k,*>=0\eqno(4.6)$$
 which implies that 
$\nu\cdot k=0$.
For $m >0$, this constraint restricts states with bosonic oscillators acting on the vacuum to satisfy $\epsilon\cdot \nu=0$.
In addition, BRST  invariance  requires that 
$$J^2_m|\phi>=0\hskip40pt (m\ge 0)\eqno(4.7)$$
This constraint restricts the possible physical states with fermionic oscillators acting on the vacuum. These states must satisfy $\epsilon\cdot\nu=0$.

There are two distinct physical spectra depending on the form of the null vector $\nu$. $\nu$ can lie entirely in the four dimensional subspace common to left and right movers or $\nu$ can have components in the internal dimensions.
We first determine the physical spectrum of the (1,2) string for compactifications
on an $E_8$ root lattice.
The vector $\nu$ can be written as a sum of a space-time and internal part as $\nu=\nu_0+\nu'$. We also use k to denote the total momentum, p for the space-time momentum, and K for the internal momentum.  
\vskip1pt 
I). Consider $\nu$ of the form: (1,0,0,1,0,...), i.e. $\nu'=0$. 
The constraints above require that $\nu \cdot k=0$.
For $k=(k_1,k_2,k_3,k_4)$, the constraint above gives a 2 dimensional momentum in a space with metric (1,1). 
The mass equation is $m^2=K^2/2+N-1/2$ in the NS sector and $m^2=K^2/2+N$ in the R sector. We have the states:
\vskip5pt
\noindent $a). b^i_{-1/2}|k,0> \hskip10pt 1\le i\le 8$
\vskip5pt
\noindent $b). |R> \hskip330pt (4.8)$ 
\vskip5pt
 There are 8 states of type (a).  In (b), there are also 8 states as can be seen from the partition function. Although the states in (b) are in the Ramond sector, they are not space-time fermions. The constraint in (4.7) allows only a subset of the usual Ramond Vacua and it is a massless scalar. The internal Ramond states form a spinor representation of SO(8).   
Note that there are not any physical states of the form $|k,K>$ if the internal bosons are compactified on an even self dual lattice. 
 In addition, there are not any states of the form $b^{\mu}_{-1/2}|k,0>$. These states are eliminated by the above constraint $k \cdot \nu =0$ with the additional constraints $\epsilon \cdot \nu$ and $\epsilon \cdot k=0$.
 In summary, when the null vector, $\nu$, has only space-time components, the states are massless scalars with momentum  in a 2 dimensional subspace with metric (1,1). Furthermore, there will be no group structure in the physical states which have a 16 fold degeneracy.
\vskip5pt 
II). Consider a null vector of the form $\nu =(1,0,0,0,1,0,...).$
For a state of the form $|k,K>$, the condition $\nu\cdot k=0$ 
gives $k_1 =K \cdot \nu'$. Since one of the components of the 
momentum is constrained, the momentum can be considered to lie in a 3 dimensional subspace with metric (1,2). The 3 dimensional mass is given by $m_3^2=-(-k_2^2+k_3^2+k_4^2)=-(K\cdot \nu')^2$. 
 We have the states:
\vskip5pt
\hskip100pt$a).\hskip3pt  |R>$ with  $k_1=0$ 
\vskip5pt
\hskip100pt $b).\hskip3pt   \epsilon_i b^i_{-1/2}|k,0>$
\vskip5pt
\hskip100pt $c). \hskip3pt  \epsilon_{\mu}b^{\mu}_{-1/2}|k,0>$ \hskip173pt (4.9)
\vskip10pt
The states in (a) are in the Ramond sector and are massless scalars because of the constraint (4.7). The space-time part is a  product of a scalar and a spinor of SO(2). So it is a space-time boson because it does not form a representation of the Lorentz group. The internal part in this case is a  product of a scalar and a spinor of S0(6). 
The constraint (4.7) applied to states in (b) requires $\epsilon \cdot \nu'=0$. This eliminates one state so there
are a total of seven states of type (b). From BRST invariance, $\epsilon \cdot \nu_0=0$, and (c) has 
only 1 solution. For this value of $\nu$, the solution for $\epsilon$ is $\epsilon=(0,1,1,1)$. The state in (c) is also a candidate state for gravity. 
In summary, in the case when the null vector $\nu$ has internal components, the states are again massless scalars, but 
the momenta is confined to a 3 dimensional subspace with metric (1,2). Furthermore, there is no group structure. The states $\alpha^i_{-1}|0>$, and  $|k,K>$ which would have group structure, are not physical.

We now investigate the theory that arises when the internal bosons are fermionized. The conformal field theories of n free bosons and 2n free fermions are equivalent in the sense that the conformal properties of their operators and all correlation functions are identical. This is true only when the bosons are compactified [10,11]. The bosons are fermionized according to: 
$$:e^{\pm iX^i(z)}:c_{\pm}={1\over \sqrt 2}(\psi^{2i-1}\pm i\psi^{2i})(z)$$
$$i\partial \phi^i(z)=:\psi^{2i}\psi^{2i-1}:(z)\eqno(4.10)$$
where $c_{\pm}$ is a cocycle which is added to ensure that the fermions
anticommute.
The internal parts of the Virasoro current, supercurrent, and u(1) current are:
$$L(z)={1\over 2}\partial_z\psi^i(z)\psi^i(z)$$
$$T_F(z)=-{i\over {12 \sqrt{k}}}f_{abc}\psi^a\psi^b\psi^c(z)$$
$$J^1(z)= -{i\over {2 \sqrt{k}}}\nu'_{a}f_{abc}:\psi^b\psi^c(z):\eqno(4.11)$$
These currents satisfy the same algebra (3.8) as the currents given in (3.5).

As in the (1,2) bosonic theory above, the physical spectrum of the fermionized theory depends on the form of the null vector $\nu$. Since we only fermionized the internal bosons, the space-time components of the U(1) current will be the same as the bosonic case. However, the states of the theory now form representations of a 24-dimensional semi-simple gauge group.
\vskip5pt 
I). Consider $\nu$= (1,0,0,1,0,...).
The mass equation is $M^2=N-1/2$ in the NS sector and $m^2=N+1$ in the R sector.
The massless states occur in the NS sector and are given by:
$$b^i_{-1/2}|k> \hskip10pt  1\le i\le 24\eqno(4.12)$$
These states could be in the adjoint representations of any 24 dimensional gauge group. Furthermore, as discussed previously, the momentum, k, lies in a 2 dimensional subspace with metric (1,1) of the 4 dimensional space. Other representations of the gauge group are possible if more sectors are added. 

II).In the case when the U(1) current contains internal components, we have states of the form $$\epsilon_{\mu}b^{\mu}_{-1/2}|k,0>\eqno(4.13)$$ This is essentially the same as in the bosonic case above and since there is only one solution for $\epsilon$, this state is a massless scalar which can describe gravity. The internal states have the form
$$\epsilon_ib^i_{-1/2}|k,0>\eqno(4.14)$$
The constraint (4.6) is the requirement that $J_0=\nu'_i f_{ibc}\sum\psi^b_{-r}\psi^c_{r}$ annihilate physical states. As before, we have the constraint (4.7). The states in (4.14) must therefore satisfy:
 
$$\nu'_if_{ibc}\epsilon_c=0 \hskip40pt \epsilon\cdot \nu'=0\eqno(4.15)$$

The U(1) current constraint breaks the original symmetry to a subset of the original 24-dimensional group. Also, as in the bosonic case, when the U(1) current contains internal components, the momentum lies in a 3 dimensional (1,2) subspace of the 4 dimensional space. However, in this theory where the internal bosons have been fermionized, there are no 3-dimensional tachyonic states. 
\vskip20pt
\centerline{\bf V. Gauge symmetry in the fermionic (1,2) heterotic string}
\vskip10pt
For the simple model with fermions in the NS and R sectors, the physical 
spectrum contains massless scalars in the adjoint representation of a 24 dimensional gauge group. It may be possible to obtain other representations of the gauge group
by adding sectors to the theory. This is one nice feature of the (1,2) string which is not present in the (0,2) string. All the gauge degrees of freedom in the (0,2) string are in the adjoint representation. It is also conceivable in the (1,2) theory to obtain
states in the $(3,2)_{Y}$ representation of the gauge group.  This is something that was not possible in the N=1 type II string because the internal space was not large enough. It has been widely believed that to obtain  SU(3)xSU(2)xU(1) in the fundamental representation, $\hat c\ge 20/3$ [12]. Our purpose here is to show that in the context of our current understanding of gauge symmetry, the $(3,2)_{Y}$ representations are possible in the (1,2) string, albeit for massless scalars rather than fermions. In the case of the heterotic (1,2) string, it is possible because the critical dimension is 12 rather than 10 and a larger symmetry group is possible.

Consider the supercurrent in the form:
$$T_F(z)=-{i\over {12 \sqrt{k}}}f_{abc}\psi^a\psi^b\psi^c+{1\over{2\sqrt{k}}}\psi^a\tilde J^a\eqno(5.1)$$
and Kac-Moody current:
$$J^a=\tilde J^a-{i\over 2}f_{abc}\psi^b\psi^c\eqno(5.2)$$ which has level
$k=\tilde k + c_A/2$ and $c_A = 2n$ for Su(n). These currents satisfy a Kac-Moody algebra:
$$[J_m^a,J_n^b]=if_{abc}J^c_{m+n}+km\delta{_ab}\delta{m+n}\eqno(5.3)$$
The generators $J_0$ give the associated Lie Algebra. 
The first term of the Kac-Moody current is the generator of the fundamental representation
and the second term is a generator of the adjoint representation. One can determine what groups are possible for a given  central charge of the Virasoro algebra[12,13].
For SU(2), we have
$$\hat c=3-{4\over {2+\tilde k}}=1,5/3,2,...$$
and for SU(3)
$$\hat c =8-{16\over {3+\tilde k}}=8/3,4,24/5,..\eqno(5.4)$$
In addition, with our current understanding, U(1) requires $\hat c$=1.   

We assume that all the group symmetry comes from the left movers, which  must be true in the (1,2) string. To obtain a state in the $(3,2)_{Y}$ representation requires that $\tilde k\ge 1$ for each group. Using the equation (5.4), we see that the contribution of SU(3) with $\tilde k \ge 1$ is $\hat c \ge 4$ and 
SU(2) with $\tilde k \ge 1$ is $\hat c \ge 5/3$. In addition, the U(1) charge can be accommodated with
$\hat c=1$. Thus, to obtain the $(3,2)_{Y}$ representations, we need $\hat c\ge 20/3$ or $c\ge 10$. In the (1,2) string we are considering, $\hat c =8$ can be achieved with $\tilde k=1$ for SU(2), and $\tilde k=3$ for SU(3). 
\vskip10pt
 
\centerline{\bf VI. One Loop Path Integral for (1,2) string}
\vskip10pt
The 1 loop path integral for the (1,2) string has the form:
$$Z=\int Dh_{\alpha\beta}DX D\psi D\psi' D\chi_{\alpha} D\bar \chi_{\alpha} D\chi_{\alpha}' DA_{\alpha}D\psi^{\prime\prime} e^{iS}\eqno(6.1)$$
where the action, S, is given in (2.3) and (2.4). For the 1-loop calculation, the path integral involves an integration over all metics of the torus, $T^2$. 
The action is invariant under reparametrizations of the world sheet coordinates(2-d general covariance), Weyl invariance, local 2-d Lorentz transformations, global target space lorentz transformations, local N=2 supersymmetry for right movers, local N=1 supersymmetry for left movers, and U(1) gauge invariance. Because the action is invariant under these symmetries, the path integral is highly divergent; one integrates infinitely many times over gauge equivalent configurations of the Action. Therefore, one must take into account the symmetries of the theory when performing the path integral.   
 Ultimately, the functional integration reduces to an integration over the fundamental region of the  moduli spaces associated with gauge inequivalent fields and a trace over the non-zero frequency modes of the oscillators associated with the fields. After considering the moduli parameters associated with the metric (teichmuller parameter), gravitini (supermoduli), and U(1) gauge field (gauge moduli), the path integral reduces to
$$Z=\int_{F}{d^2\tau\over \tau_2 vol(ckv)}\int {d^2\hat A \tau_2\over vol(cks)}(czero)(t zero)\int{d^4p}e^{-2\pi p^2\tau_2}gtr'g'q^{L_0}\bar q^{\bar L_0}\eqno(6.2)$$
where the prime denotes the omission of the zero frequency modes, $q=e^{2\pi i \tau}$,$\tau=\tau_1+i\tau_2$ is the complex metric moduli parameter,  and g is a twist factor. In the sectors where no A-field modulus is present, g=1 for time boundary conditions which are periodic on bosons and antiperiodic on fermions, and $g=(-1)^F$ for time boundary conditions which are
antiperiodic on bosons and periodic on fermions. Here, F is the number operator for the relevant oscillator. When there is an A-field modulus, g contains an additional factor of $w^{J_0}$ where $w= e^{i\tau_2 A_z}$ and $J_0$ is the generator of U(1) transformations. $g^{\prime}$ contains the part of the U(1) current without momentum. The volume of the conformal killing vectors, vol(ckv), is $\tau_2^2$; the volume of the conformal killing scalars, vol(cks), is $4\pi^2\tau_2$; the c zero modes contribute a factor of $\tau_2^2$; and the U(1) ghost zero modes t,  contribute a factor of $\tau_2^2$. The analysis here parallels that of the calculation of the 1-loop amplitude for N=2 strings given in [9]. 

\noindent Each contribution to the trace is of the form:

$$\Pi_{n=Z-r}^{\infty}(1\pm wq^n)^{\pm 1}\eqno(6.3)$$

The powers $\pm 1$ occur for fermions and bosons respectively, the factor w is absent if there is no gauge-field modulus, and the relative sign depends on time boundary conditions: + for antiperiodic fermions or periodic bosons and - for
periodic fermions or antiperiodic bosons. These can also be expressed in terms of Jacobi theta functions(see appendix). 
The contribution from the right moving N=2 part is 1 regardless of the spin structure. The space-time part and the ghosts of the left moving part also cancel for all spin structures( Note, however, that the normal ordering constant does depend on the boundary conditions of the space-time fields).

The Wilson loop (or g factors) for the left movers  shifts the momentum $k^{\mu} \rightarrow k^{\mu}+{A_z\tau_2\over{4\pi\tau}} \nu^{\mu}$.
Including the wilson loop would mean that one integrates over the real momenta an infinite number of times. It also includes complex momentum.  We restrict the integration by identifying the momentum with its spectral flow, i.e., we drop the wilson loop for the left movers.  The integration over the wilson loop for the right moving N=2 part is equivalent to summing over all fermionic boundary conditions. In this case, we restrict the integration to include each boundary condition only once by chosing a fundamental region of the gauge moduli space. The volume of this moduli space is given by $\int d^2 \hat A={4\pi^2\over {\tau_2}} $. Considering all these factors, the partition function becomes:

$$Z={1\over 4}\int_{F}{d^2\tau \over {\tau^2_2}} \sum_{proj}Tr^{\prime}q^{ L_{0}}(-1)^{F_p}\eqno(6.4)$$

\noindent The trace over the nonzero frequency modes of the oscillators can be written as:
$$Tr^{\prime}q^{ L_{0}}(-1)^{F_p}=
q^{-a_{p}}Tr^{\prime}q^{N_{st}}Tr^{\prime}q^{N_{g}}Tr^{\prime}q^{N_{int}}\sum q^{1/2 K^2}(-1)^{F_p}\eqno(6.5)$$
Here we have used the notation that $N_{st}$ is the space-time part of $\tilde L_0$, $N_g$ is the ghost part, and $N_{int}$ is the internal part. 
  
 We consider two sectors: one sector has NS boundary conditions for all 12 left moving fermions  and another sector has R boundary conditions for all left moving fermions. We then add in projections in order to include all boundary conditions for both directions on the torus.  The integration over the gauge moduli parameters is equivalent to summing over all boundary conditions for the right moving fermions(N=2 part). 

\noindent NS sector:(a=1/2)
$$A(-,-) =\eta(-,-)q^{-1/2}Tr^{\prime}q^{N_b}\sum q^{1/2 K^2}Tr^{\prime}q^{N_{\alpha}}=\eta(-,-){{\theta_3^4}\over {\eta ^{12}}}\sum q^{1/2 K^2}$$
$$A(-,+)=\eta(-,+)q^{-1/2}Tr^{\prime}q^{N_b}Tr^{\prime}q^{N_{\alpha}}(-1)^F\sum q^{1/2 K^2}=\eta(-,+){{\theta_4^4}\over {\eta ^{12}}}\sum q^{1/2 K^2}$$
\noindent R sector(a=0)
$$A(+,-)=\eta(+,-)Tr^{\prime}q^{N_d}Tr^{\prime}q^{N_{\alpha}}\sum q^{1/2 K^2}=\eta(+,+)2^{-8/2}{{\theta_2^4}\over {\eta ^{12}}}\sum  q^{1/2 K^2}$$
$$A(+,+)=\eta(+,+)Tr q^{N_d}(-1)^FTr^{\prime}q^{N_{\alpha}}\sum q^{1/2 K^2}=0\eqno(6.6)$$
The constants $\eta$ are determined by requiring the partition function 
to be modular invariant. Only the relative phases of the projections are relevant, so we can set $\eta(-,-)=1$. We then determine that $\eta(-,+)=-1$ and
$\eta(+,-)=-2^4$.
Compactifying the internal bosons on the $E_8$ lattice, we find that the  partition function is 
$$Z={1\over 4}\int_F {d^2\tau \over {\tau_2^2}}{1\over 2}{{[\theta_3^8+\theta_4^8+\theta_2^8]}\over {\eta^8}}[({\theta_3\over {\eta}})^4-({\theta_4\over {\eta}})^4-({\theta_2\over {\eta}})^4]\eqno(6.7)$$

This one-loop 0 point amplitude is modular invariant, i.e. it does not change under changes in the metric moduli parameter $\tau'=\tau + 1$, and $\tau'=-{1\over \tau}$. It is also 0 because of the Hardy Ramanujan identity.  
\vskip10pt
We now consider the same model but now we fermionize the internal bosons. 
We still have two sectors: one sector with  NS boundary conditions on all 28 left moving fermions and another sector with R boundary conditions for all 28 left moving fermions. We include all spin structures on the torus by adding in projections. The space-time parts are the same as before. 
 
NS sector:(for 24 internal fermions+ 4 from space-time, a=1/2)
$$A(-,-) =\eta(-,-)q^{-1/2}Tr^{\prime}q^{N_b}=\eta(-,-){{\theta_3^{12}}\over {\eta ^{12}}}$$
$$A(-,+)=\eta(-,+)q^{-1/2}Tr^{\prime}q^{N_b}(-1)^F=\eta(-,+){{\theta_4^{12}}\over {\eta ^{12}}}$$
R sector(a= -1)
$$A(+,-)=\eta(+,-)q Tr^{\prime}q^{N_d}=\eta(+,-)2^{-24/2}{{\theta_2^{12}}\over {\eta ^{12}}}$$
$$A(+,+)=\eta(+,+)q Tr q^{N_d}(-1)^F=0\eqno(6.8)$$
Here, $N_b$ is the part of $\tilde L_0$ containing fermions with NS boundary conditions, and $N_d$ is the part containing fermions with Ramond boundary conditions. Setting $\eta(-,-)=1$, modular invariance requires $\eta(-,+)=-1$, and $\eta(+,-)=-2^{12}$. Finally, the partition function is:
$$Z={1\over 4}\int_F {d^2\tau \over {\tau_2^2}}[({\theta_3\over {\eta}})^{12}-({\theta_4\over {\eta}})^{12}-({\theta_2\over {\eta}})^{12}]\eqno(6.9)$$

\vskip20pt
\centerline{\bf VII.  Effective field theory}
\vskip10pt

We can determine an effective field theory for the (1,2) string theory. This method involves calculating scattering 
amplitudes for the string theory and then determining an action with space-time fields which reproduces the string scattering amplitudes.
We concentrate on the two dimensional case where the only coupling involves the massless scalars.

The vertex operator for the massless scalars in the adjoint representation of the gauge group corresponding to the state, $b^a_{-1/2}|k>$, is :
$V^a=V^a_LV_R$ where
$$V^a_L=\{{k\cdot b(z)b^a(z)-{i\over 2}f_{abc}b^b(z)b^c(z)}\}exp\{{ik\cdot X_L(z)}\}\eqno(7.1a)$$
$$V_R=\{{ik\cdot \bar \partial \bar X_R(\bar z)-i\bar k\cdot \bar \partial X_R(\bar z)-(k\cdot \bar \psi_R(\bar z))(\bar k\cdot \psi_R(\bar z))}\}exp\{{ik\cdot \bar X_R(\bar z)+i\bar k\cdot X_R(\bar z)}\}\eqno(7.1b)$$

The left moving part (7.1a) is expressed in terms of real momentum and the right moving part (7.1b) is expressed in terms of complex momentum. 
In the two dimensional case, the complex momentum of the physical states reduces to 2 real components. 
The three point scattering amplitude for these massless scalars is:
$$A_3=<-k_1|b^a_{1/2}V^b(k_2;z,\bar z=1)b^c_{-1/2}|k_3>=(k_1\cdot \bar k_2-\bar k_1\cdot k_2)i f_{abc}(2\pi)^4\eqno(7.2)$$
An effective action which reproduces this three point coupling is:
$$S=\int d^4x\big(1/2\partial_i\phi^a\bar \partial_i\phi^a+{1\over 3!}f^{abc}\phi^a\partial_i\phi^b\bar \partial_i\phi^c+O(\phi^4)\big)\eqno(7.3)$$
This effective action will give rise to higher n point scattering amplitudes. The scattering amplitudes for the massless scalars of the (1,2) string are the same as that for the (0,2) string[5].Since the higher n point amplitudes derived from the string theory are zero, one must add higher order terms in the action so that the total contribution to the four-point coupling vanishes.  The effective action which reproduces the scattering amplitudes of the two- dimensional theory essentially describes a scalar theory of self-dual Yang Mills. This is given by [5]:
$$S=\int k^{(0)}\wedge Tr\big(1/2\bar\partial\phi\wedge\partial\phi+\sum_{n=3}^{\infty}{1\over n!}\bar \partial \phi\wedge[...[\partial\phi,\phi],\phi]...\big)\eqno(7.4)$$ 
$k^{(0)}$ is the kahler form, $k^{(0)}=i(dx^1\wedge d\bar x^1-dx^2\wedge d\bar x^2)$.
The action for self-dual Yang Mills in terms of scalars is derived by solving the self dual equation $F_{\mu \nu}=\epsilon_{\mu\nu\rho\sigma}F^{\rho\sigma}$. The gauge field can be written in terms of scalar fields, $A_{\mu}=i\partial_{\mu}\Omega^{-1}\Omega$ and $\Omega=exp[\sum_a\phi^at^a]$
,  where $t^a$ is a hermitian generator of the gauge group, and $\phi^a$ is a scalar field with the adjoint index. Substituting these expressions for the gauge field into the self-duality equation results in a non-linear differential equation for the scalar fields. This equation can be derived from the action (7.4).

In the 3 dimensional case, the gauge degrees of freedom couple to gravity. This will modify the two dimensional theory and the theory is no longer 
described by self dual Yang Mills.

\vskip10pt
 
\vskip20pt
\centerline{\bf VIII. Conclusions}
\vskip10pt
The heterotic N=2 theories have a finite number of physical particles in the
spectrum, all of which are massless, and occur in two or three space-time dimensions. This paper concerned the (1,2) 
heterotic string, which is constructed with a combination of  a string with local N=1 worldsheet supersymmetry for the left movers and a local N=2 worldsheet supersymmetry for the right movers. An action describing this theory is given in sect. II. This theory exists in either two or three space-time dimensions depending on the gauged U(1) left moving current. When the  vector $\nu$ describing the U(1) current has no internal components, the physical spectrum consists of a number of massless scalars propagating in two dimensional space-time with metric (1,1). When the  vector $\nu$ has internal components, the spectrum  consists of a number of scalars, some massless and some tachyonic in 3 space-time dimensions with signature (1,2). In this case, there is one state that is a candidate to describe gravity. The (1,2) string  compactified on an even self-dual lattice has massless states but no gauge symmetry. Fermionizing the internal bosons and choosing a supercurrent trilinear in fermions allows for non-abelian gauge symmetry. The physical spectrum again depends on whether the  vector $\nu$ has internal components. When the  vector does not have internal components, the physical spectrum consists of a number of massless scalars which form a representation of any 24-dimensional semi simple gauge group. Unlike the (0,2) string, the (1,2) string can accommodate fundamental representations of a gauge group. The internal space is also large enough to accommodate particles in the fundamental representations of SU(3)xSU(2)xU(1).
This two-dimensional version has only gauge degrees of freedom, and its effective action is equivalent to self-dual Yang Mills. Thus, these states quantize self-dual Yang Mills. 
When the vector $\nu$ has an internal component, the physical spectrum consists of a number of scalars in 3 space-time dimensions. The group symmetry in this case is broken to a subset of the original 24-dimensional gauge group. In this three dimensional theory, the coupling of the gauge fields to gravity modifies the effective action and the theory is no longer described by self-dual Yang Mills.  
 The partition function is calculated for a (1,2) theory compactified on an $E_8$ lattice and for a theory where the internal bosons are fermionized. By summing over spin structures of the left moving fermions and choosing suitable projections, we derived a modular invariant partition function. 

\vskip20pt

\centerline{\bf Appendix}
\vskip10pt

Modular invariance of the partition function is  more easily seen if the traces of the fields
are expressed in terms of Jacobi theta funcions. 

$$Tr^{\prime}q^{\sum\alpha^i_{-n}\alpha^i_{n}}=f(q)^{-d}=q^{d\over 24}\eta^{-d}\eqno(A.1)$$
$$Tr^{\prime}q^{\sum rb_{-r}b_{r}}=({\theta_3\over \eta})q^{d_b\over 48}\eqno(A.2)$$
$$Tr^{\prime}q^{\sum rb_{-r}b_{r}}(-1)^F=({\theta_4\over \eta})q^{d_b\over 48}\eqno(A.3)$$
$$Tr^{\prime}q^{\sum rd_{-r}d_{r}}(-1)^F=(2^{-d/2})({\theta_2\over \eta})q^{-2d_d\over 48}\eqno(A.4)$$
$E_8$ root lattice:
$$\sum q^{1/2 p_i^2}=1/2(\theta^8_3+\theta^8_4+\theta^8_2)\eqno(A.5)$$
The Jacobi theta functions are given by:
$$\theta_2=\vartheta\left[{1\atop 0}\right](0|\tau)$$
$$\theta_3=\vartheta\left[{0\atop 0}\right](0|\tau)$$
$$\theta_4=\vartheta\left[{0\atop 1}\right](0|\tau)\eqno(A.6)$$
where 
$$\vartheta\left[{\rho\atop\mu}\right](\nu|\tau)=\sum_{n\in Z}e^{i\pi\tau(n+\rho/2)^2}e^{i2\pi(n+\rho/2)(\nu+\mu/2)}e^{i\pi\rho\mu/2}\eqno(A.7)$$
$$\eta(q)=q^{1/24}\Pi_{n=1}^{\infty} (1-q^n)\eqno(A.8)$$

\vskip20pt

\centerline{\bf References}

\vskip10pt
\item{1.} M. Ademollo, L. Brink, A. D'Adda, R. D'auria, E. Napolitano, S. Sciuto, E. Del Giudice, P. DiVecchia, S. Ferrara, F. Gliozzi, R. Musto, R. Pettorino, J. H. Schwarz, Nucl. Phys. B111 (1976) 77
\item{2.} A. D'Adda and F. Lizzi, Phys. Lett. B191(1987)85
\item{3.} E.S. Fradkin and A.A. Tseytlin, Phys. Lett. B106 (1981) 63;
          S. Mathur and S. Mukhi, Phys. Rev. D36 (1987)465
          S. Ketov, O. Lechtenfeld, Phys. Lett. 353B(95)373
\item{4.} M. Green, Nucl. Phys. B293 (1987) 593:           
\item{5.} H. Ooguri and C. Vafa, Nucl. Phys B367 (1994)83
\item{6.} I.B. Frenkel, H.Garland, E.J. Zuckerman, Proc. Nat. Acad. Sci., 83(1986)8442
\item{7.} W. Siegel, Nucl. Phys. B238(1984)307
\item{8.} E. Fradkin, G. Vilkovsky, Phys. Lett. B55 (1975)224
\item{9.} S. Mathur, S. Mukhi, Nucl. Phys. B302 (1988)130
\item{10} S. Coleman, Phys. Rev. D11 (1975) 2088
\item{11} S. Mandelstam, Phys. Rev. D11 (1975) 3026; M. B. Halpern, Phys. Rev. D4 (1971)2398
\item{12} L.J. Dixon and V. Kaplunovsky, C. Vafa, NP B294(1987) 43
\item{13} J. Schwarz in Superstrings, eds. G.O. Freund and  K.T. Mahanthappa \vskip1pt (Plenum Press 1988), p. 117
     
\vskip20pt
\centerline{\bf Acknowledgements}

The author would like to thank Louise Dolan for many helpful discussions.

\noindent This work is supported in part by the US Department of Energy under grant
DE-FG-05-85ER-40219/Task A.
\vskip10pt    

\end